# Impact of Fe site Co substitution on superconductivity of $Fe_{1-x}Co_xSe_{0.5}Te_{0.5}$ (x = 0.0 to 0.10): A flux free single crystal study

P.K. Maheshwari[1,2], Bhasker Gahtori[1], Anurag Gupta[1,2] and V.P.S. Awana[1,*]

[1]CSIR-National Physical Laboratory, Dr. K. S. Krishnan Marg, New Delhi-110012, India

[2]Academy of Scientific and Innovative Research, NPL, New Delhi-110012, India

**Abstract**

We report synthesis of Co substitution at Fe site in $Fe_{1-x}Co_xSe_{0.5}Te_{0.5}$ (x=0.0 to 0.10) single crystals via vacuum shield solid state reaction route using flux free method. Single crystal XRD results showed that these crystals grow in (00*l*) plane i.e., orientation in *c*-direction. All the crystals possess tetragonal structure having *P*4/*nmm* space group. Detailed scanning electron microscopy (SEM) images show that the crystals are grown in slab-like morphology. The EDAX results revealed the final elemental composition to be near stoichiometric. Powder X-Ray diffraction (PXRD) Rietveld analysis results show that (00*l*) peaks are shifted towards higher angle with increasing Co concentration. Both *a* and *c* lattice parameters decrease with increasing Co concentration in $Fe_{1-x}Co_xSe_{0.5}Te_{0.5}$ (x=0.0 to 0.10) single crystals. Low temperature transport and magnetic measurements show that the superconducting transition temperature ($T_c$), decreases from around 12K to 10K and 4K for x=0.03 and x=0.05 respectively. For x=0.10 crystal superconductivity is not observed down to 2K. Electrical resistivity measurement of $Fe_{0.97}Co_{0.03}Se_{0.5}Te_{0.5}$ single crystal under magnetic field up to 14Tesla for *H//ab* and *H//c* clearly showed the anisotropy nature of superconductivity in these crystals. The upper critical field $H_{c2}(0)$, being calculated using conventional one band Werthamer–Helfand–Hohenberg (WHH) equation, for x=0.03 crystal comes around 70Tesla, 45Tesla and 35Tesla for normal state resistivity criterion $\rho_n$= 90%, 50% and 10% criterion respectively for *H//c* and around 100Tesla, 75Tesla and 60Tesla respectively for *H//ab*. The activation energy of $Fe_{0.97}Co_{0.03}Se_{0.5}Te_{0.5}$ single crystal is calculated with the help of TAFF model for both *H//c* and *H//ab* direction. In conclusion, Co substitution at Fe site in $Fe_{1-x}Co_xSe_{0.5}Te_{0.5}$ suppresses superconductivity.

Key words: Fe chalcogenide superconductivity, Co doping, Structural details, Magneto transport and Magnetization.

PACS number(s): 74.70.-b, 74.25.F-, 74.10.+v, 74.25.Ha

*Corresponding Author

Dr. V. P. S. Awana: E-mail: awana@mail.nplindia.org

Ph. +91-11-45609357, Fax-+91-11-45609310

Homepage awanavps.webs.com

**Introduction**

Discovery of superconductivity in Fe based superconductor had been one of the most surprising discoveries for both experimental and theoretical condensed matter physicist [1-6]. Both iron pnictides [1-4] and chalcogenides [5, 6] based superconductors had become one of the biggest surprises after the discovery of high $T_c$ (*HTSc*) cuprates [7, 8] in condensed matter physics. The ground state of parent non superconducting Fe based compounds is known to be magnetically ordered one. Fe based superconducting compounds i.e., pnictides and chalcogenides are also known to be outside the well known conventional superconducting BCS theory [9], similar to high $T_c$ cuprates. For the condensed matter theorist's community, theoretical explanation for high $T_c$ superconductor is biggest challenge yet.

As far as iron chalcogenides [Fe(Se$_{1-x}$Te$_x$)] are concerned, they possess simplest crystal structure. Though FeTe does not show any superconducting transition [10], the FeSe exhibits superconducting transition at around 8K [11]. In FeSe$_{1-x}$Te$_x$ series, the highest $T_c$ of around 15K is seen for the FeSe$_{0.5}$Te$_{0.5}$ at ambient pressure. [12-14]. 15K superconductivity of FeSe$_{0.5}$Te$_{0.5}$ is reported to decrease with 3d metal (Ni/Co/Cr/Zn) doping at Fe site [15-19]. For in depth understanding of the effect of 3d metal doping at Fe site in FeSe$_{0.5}$Te$_{0.5}$, one would prefer doped single crystals rather than poly-crystal ones. Theoreticians would like to model single crystals data rather than the poly-crystals, as the latter ones do have extrinsic (grain boundary etc.) contributions along with the intrinsic material characteristics. In case of Fe chalcogenide superconductors, the most of the literature reported pertaining to Fe site 3d metal doping is on polycrystalline samples [15-18]. The single crystal growth of Fe chalcogenide superconductors is quite complicated and is mostly possible with added flux (KCl/NaCl) only, further the obtained single crystal were of tiny (mm) size [20-23]. The flux free growth of Fe chalcogenides single crystal is possible by using Bridgman method with complicated heating schedules [24-26]. However, the flux free growth of large single crystals of FeSe$_{1-x}$Te$_x$ (x=0.00 and x=0.50) was reported recently [27-29] with simple heat treatment schedule and on normal automated furnace [28, 29]. Very recently, the flux free growth of Fe site Ni doped Fe$_{1-x}$Ni$_x$Se$_{0.5}$Te$_{0.5}$ single crystals was reported by us and fast suppression of superconducting transition temperature was seen with Ni content [30].

Keeping in view, the need of flux free 3d metal doped FeSe$_{0.5}$Te$_{0.5}$ large single crystals and in continuation to our recent work on Fe$_{1-x}$Ni$_x$Se$_{0.5}$Te$_{0.5}$ single crystals [30], here we report flux free large single crystal (cm size) growth and superconductivity properties of Fe site Co doped Fe$_{1-x}$Co$_x$Se$_{0.5}$Te$_{0.5}$ (x=0.0 to 0.10) series, with the similar protocol reported very recently for Fe$_{1-x}$Ni$_x$Se$_{0.5}$Te$_{0.5}$ [30]. Large single crystals of Fe$_{1-x}$Co$_x$Se$_{0.5}$Te$_{0.5}$ are obtained till 10% Co doping at Fe site. It is seen that Co substitution at Fe site in $Fe_{1-x}Co_xSe_{0.5}Te_{0.5}$ suppresses superconductivity, although the rate of $T_c$ suppression is much less in comparison to $Fe_{1-x}Ni_xSe_{0.5}Te_{0.5}$ [30]. In our knowledge, this is first detailed study on $Fe_{1-x}Co_xSe_{0.5}Te_{0.5}$ (x=0.0 to 0.10) series of large flux free single crystals, earlier studies were mostly either on tiny and flux assisted single crystals [20-23] or on polycrystalline samples [15-18].

**Experimental Details**

All the $Fe_{1-x}Co_xSe_{0.5}Te_{0.5}$ (x=0.0 to 0.10) single crystals were grown in a normal automated furnace using flux free method. The essential elements i.e., Fe, Co, Se and Te powder of high purity better than 4N are taken in stoichiometric ratio i.e., $Fe_{1-x}Co_xSe_{0.5}Te_{0.5}$ (x=0.0 to 0.10) and grinded in Ar gas filled glove box. This grinded powder is pelletized by applying hydrostatic pressure of 100kg/cm$^2$, and sealed in a quartz tube with the vacuum of less than 10$^{-3}$ torr. These sealed quartz tubes are kept in a normal programmable automated furnace, and heated up to 1000$^0$C for 24 hours with an intermediate step of 450$^0$C with the rate of 2$^0$C/minute. Finally the furnace is cool down very slowly up to room temperature with the rate of 10$^0$C/hour. Thus obtained crystals were very shiny and big in size and their photographs are shown in Figure 1. X-ray diffraction (*XRD*) of as obtained single crystals and their crushed powder was done using Rigaku X-ray diffractometer with CuKα radiation of 1.54184Å at room temperature. Scanning electron microscopy (SEM) pictures has been taken on ZEISS-EVO-10 electron microscope to understand the morphology of $Fe_{1-x}Co_xSe_{0.5}Te_{0.5}$ single crystals. Electrical and magnetic measurements were performed on Quantum Design Physical Property Measurement System (PPMS-14Tesla) down to 2K and up to magnetic field of 14Tesla.

**Results and Discussion**

Figure 1 shows the photograph of a piece of as synthesized $Fe_{1-x}Co_xSe_{0.5}Te_{0.5}$ (x=0.0 to 0.10) single crystals after breaking the quartz tube. All the synthesized samples are looking very shiny and in single crystalline form. As synthesized all the crystals are of cm sizes. As all the samples are synthesized via self flux method so there is no requirement to remove foreign flux. The large size flux free method grown crystals are certainly more attractive for the physical property measurements. All the measurements were done on a small piece, taken from obtained crystals.

SEM is very useful tool to understand the morphology of the as synthesized material. Figure 2 (a-d) shows the SEM and EDAX results of the $Fe_{1-x}Co_xSe_{0.5}Te_{0.5}$ for x=0.03 and 0.10 crystals.

Un-doped x=0.00 i.e., FeSe$_{0.5}$Te$_{0.5}$ single crystal SEM and EDAX results are already shown elsewhere [28]. Fig 2 (a) and 2(b) show the SEM images of Fe$_{0.97}$Co$_{0.03}$Se$_{0.5}$Te$_{0.5}$ and Fe$_{0.90}$Co$_{0.10}$Se$_{0.5}$Te$_{0.5}$ single crystal respectively. It is clearly seen that the morphology of the synthesized crystals is slab like and the growth is layer by layer. The morphology of these crystals is similar to FeSe$_{1-x}$Te$_x$ single crystals being reported earlier [28-30]. EDAX analysis is used for compositional analysis of the studied crystals. Fig 2(c) shows the quantitative analysis of selected area of Fe$_{0.97}$Co$_{0.03}$Se$_{0.5}$Te$_{0.5}$ single crystal. The result of compositional analysis is found to be in stoichiometric ratio i.e., close to Fe$_{0.97}$Co$_{0.03}$Se$_{0.5}$Te$_{0.5}$ with only a slight loss of Se. Fig 2 (d) depicts analysis of the selected area for Fe$_{0.90}$Co$_{0.10}$Se$_{0.5}$Te$_{0.5}$, showing all the elements in near stoichiometric ratio without presence of impurity elements. It is clear from SEM results shown in Fig. 2 that the studied Fe$_{1-x}$Co$_x$Se$_{0.5}$Te$_{0.5}$ crystals are slab like and their growth is layer by layer. Further they are near stoichiometric, i.e., the starting composition is intact.

Fig 3 shows the single crystal XRD pattern of Fe$_{1-x}$Co$_x$Se$_{0.5}$Te$_{0.5}$ (x=0.0 to 0.10) crystals at room temperature. It is clearly seen from Fig 3 that growth of all the crystal is in (00*l*) plane, as the high peak intensities appeared only in [001],[002] and [003] planes. These results confirm the crystalline nature of the studied samples. Further, the XRD results show that (00*l*) peaks are shifted towards higher angle with increasing Co doping in Fe$_{1-x}$Co$_x$Se$_{0.5}$Te$_{0.5}$ single crystals. This is the indication of decrement in *c* lattice parameter with Co concentration in Fe$_{1-x}$Co$_x$Se$_{0.5}$Te$_{0.5}$. This result concurs with earlier results on Co doped FeSe$_{0.5}$Te$_{0.5}$ polycrystalline samples [18]. For more understanding about the structural property like lattice parameters, co-ordinate positions etc., we gently crushed these crystals into powder form and then performed powder XRD at room temperature. The resultant powder XRD of all crushed Fe$_{1-x}$Co$_x$Se$_{0.5}$Te$_{0.5}$ crystals is shown in Fig 4. All the samples are crystallized in tetragonal structure having *P4/nmm* space group. No other phase present in these samples are observed from powder XRD result. For further analysis we performed Rietveld refinement on Fe$_{1-x}$Co$_x$Se$_{0.5}$Te$_{0.5}$ (x=0.0 to 0.10) using Fullprof suite software of the observed powder XRD pattern. Co-ordinate position for different *z* is refined for Se and Te atom at 2*c* site. Rietveld refinements results are given in Table 1. These results show a monotonically decrement in both *a* and *c* lattice parameters with increasing Co concentration at Fe site in Fe$_{1-x}$Co$_x$Se$_{0.5}$Te$_{0.5}$ single crystals. Decrement in *c* lattice parameter is more in comparison to *a* lattice parameter. Both lattice parameters *a* and *c* decrease from 3.80Å to 3.78Å and 5.99Å to 5.93Å respectively for x=0.00 to x=0.10. For further clarification, inset view of Fig 4 i.e., zoomed part of powder XRD clearly shows the [003] peak shifts towards the higher angle, clearly representing the decrement in *c* lattice parameter.

It is clear from the structural analysis that all the studied crystals of $Fe_{1-x}Co_xSe_{0.5}Te_{0.5}$ (x=0.0 to 0.10) are cm size (Fig.1), with slab like morphology (Fig 2) and their growth is in single oriented direction i.e., in (00$l$) plane (Fig 3). Powder XRD and Rietveld refinement results (Fig. 4) show shrinkage in volume, as both *a* and *c* lattice parameter decrease, with Co concentration in $Fe_{1-x}Co_xSe_{0.5}Te_{0.5}$ (x=0.0 to 0.10) crystals. Thus confirming that smaller ion Co is substituted successfully at Fe site in $Fe_{1-x}Co_xSe_{0.5}Te_{0.5}$ till x = 0.10.

The resistivity verses temperature ($\rho$-$T$) plots for all the $Fe_{1-x}Co_xSe_{0.5}Te_{0.5}$ (x=0.0 to 0.10) single crystals in temperature range of 300K to 2K at zero magnetic field are shown in Fig 5. The $\rho$-$T$ plots are normalized with resistivity value at 300K. From Fig 5 it can be seen that there is some curvature in $\rho$-$T$ plots, suggesting change in metallic behavior while cooling down the temperature from 300K. For x=0.00, the $\rho$-$T$ plot curvature occurs at around 150K, exhibiting metallic behavior below this temperature. Further, x =0.0 crystal exhibits superconductivity with onset $T_c$ at 14K and $T_c(\rho=0)$ at 12K. In case of x=0.03 the metallic behavior starts form approximately 100K and the same becomes superconducting with $T_c$ (onset) at 12K and $T_c(\rho=0)$ at 10K. For x=0.05, the $\rho$-$T$ behavior changes to semiconducting at 80K and becomes superconducting with $T_c$ (onset) at 8K and $T_c(\rho=0)$ at 4K. In case of x=0.10 crystal the $\rho$-$T$ behavior is altogether semiconducting and the same does neither shows $T_c$ (onset) nor the $T_c(\rho=0)$ down to 2K. Clearly Co substitution at Fe site in $Fe_{1-x}Co_xSe_{0.5}Te_{0.5}$ changes normal state conduction from metallic to semi-metallic and suppresses the superconductivity. Both ensuing disorder due to $Fe^{3+}$ site $Co^{2+}$ substitution and decreasing carrier density could be the reason behind the same. Inset view of Fig 5 shows the zoomed part of Fig 5 in temperature range from 20K to 2K, clearly depicting the $T_c$ values for various Co doped crystals. Though clearly the $T_c$ decreases with Co substitution at Fe site in $Fe_{1-x}Co_xSe_{0.5}Te_{0.5}$ (x=0.0 to 0.10), the rate of suppression is not as fast as in case of Ni doping at Fe site in $Fe_{1-x}Ni_xSe_{0.5}Te_{0.5}$ single crystals [30]. For example, though superconductivity is completely suppressed for x = 0.07 in $Fe_{1-x}Ni_xSe_{0.5}Te_{0.5}$ and semiconducting behavior was seen down to 2K [30], but the same happens only at x =0.10 for studied $Fe_{1-x}Co_xSe_{0.5}Te_{0.5}$.

Since the Fe chalcogenides [$Fe(Se_{1-x}Te_x)$] superconductors are layered and their growth is in *c*-direction (00$l$) plane, one would like to check the superconductivity critical parameters in both in plane and out of plane directions. This will serve as a check for the crystalline (perfect orientation) nature of the studied crystals. For verification of the anisotropy of the superconducting properties, we measured the electrical resistivity under various magnetic fields in *H//ab* and *H//c* situations for the $Fe_{0.97}Co_{0.03}Se_{0.5}Te_{0.5}$ single crystal with the help of Quantum Design rotator. The results of this measurement are shown in Fig 6(a) and Fig 6(b) for *H//c* and *H//ab* respectively. Resistivity versus temperature behavior is different with magnetic field in both *H//c* and *H//ab*, and clearly confirms the anisotropic behavior of the crystal with magnetic fields. Fig 6(a) shows the temperature dependent electrical resistivity of $Fe_{0.97}Co_{0.03}Se_{0.5}Te_{0.5}$ single crystal with magnetic field up to 12Tesla for *H//c*. For *H//c*, the $T_c(\rho=0)$ value decreases from around 10K to 6K with increasing magnetic field from 0 to 12Tesla. The shape and broadening of *ρ-T* plots can be explained with the help of vortex liquid state which is similar to reported literature on Fe chalcogenides superconductors [20-30]. Fig 6(b) shows the *ρ-T* measurement of $Fe_{0.97}Co_{0.03}Se_{0.5}Te_{0.5}$ single crystal with magnetic field up to 14Tesla for *H//ab* situation. The $T_c(\rho=0)$ for *H//ab* decreases from around 10K to 7.5K with increasing magnetic field up to 14Tesla. The transition temperature width of *H//c* is much wider than *H//ab* situation in *ρ-T* measurement with magnetic field. These results clearly shows that suppression of superconductivity with applied magnetic field in *H//c* situation is faster than in *H//ab*. This further ascertains that the crystal growth is unidirectional i.e., in (00*l*) plane, as the magnetic field suppress superconductivity faster when applied magnetic field is perpendicular to the direction of the crystal growth plane. The superconducting transition temperature width difference of approximately 1.5K up to 12Tesla magnetic field for in plane and out plane field directions shows the presence of critical field anisotropy in $Fe_{0.97}Co_{0.03}Se_{0.5}Te_{0.5}$ single crystal.

Upper critical field ($H_{c2}$) plays an important role to understand the effect of magnetic field on superconductivity. To determine the upper critical field, the normal resistivity criteria ($\rho_n$) of 90%, 50% and 10% are used in both *H//c* and *H//ab* situations for $Fe_{0.97}Co_{0.03}Se_{0.5}Te_{0.5}$ single crystal. Fig 7(a) and Fig 7(b) show the plots between $H_{c2}$ and temperature for *H//c* and *H//ab* situations respectively. The upper critical field $H_{c2}(0)$ at zero temperature is calculated by applying the conventional one-band Werthamer-Helfand-Hohenberg (WHH) equation, i.e., $H_{c2}(0) = -0.693T_c(dH_{c2}/dT)_{T=Tc}$. In both Fig 7(a) and Fig 7(b), the solid lines represents the extrapolation of the Ginzburg–Landau equation i.e. $H_{c2}(T) = H_{c2}(0)(1-t^2/1 + t^2)$, where t = $T/T_c$ is define as reduced temperature. Calculated upper critical field $H_{c2}(0)$ with different normal state resistivity criterion of $\rho_n$= 90%, 50% and 10% is around 70Tesla, 45Tesla and 35Tesla respectively for *H//c*, and is plotted in Fig 7(a). While for *H//ab* normal resistivity criterion of $\rho_n$= 90%, 50% and 10%, the $H_{c2}(0)$ is around 100Tesla, 70Tesla and 60Tesla respectively, which is plotted in Fig 7(b). Clearly, the $H_{c2}(0)$ values in *H//ab* situation are much higher than in *H//c*. Also, when compared with un doped pristine $FeSe_{0.5}Te_{0.5}$ single crystal [28], the obtained values of $H_{c2}(0)$ for $Fe_{0.97}Co_{0.03}Se_{0.5}Te_{0.5}$ are less in both *H//ab* and *H//c* conditions. Importantly, calculated $H_{c2}(0)$ values for both in plane and out plane far exceed the Pauli Paramagnetic limit of $1.84T_c$. This is the indication of heavy pinning or exotic nature of superconductivity in $Fe_{0.97}Co_{0.03}Se_{0.5}Te_{0.5}$ crystals.

The Ginzburg–Landau coherence length $\xi(0)$ is calculated using the value of $H_{c2}(0)$ for both $H//c$ and $H//ab$. The relation between coherence length $\xi(0)$ and $H_{c2}(0)$ is given by the equation $H_{c2}(0)=\varphi_0/2\Pi\,\xi(0)^2$, where $\varphi_0$ is called flux quantum whose value is 2.0678 X $10^{-15}$ Tesla-m$^2$. At zero temperature, the coherence length $\xi(0)$ is calculated to be 21.6Å for $H//c$ and 18.14Å for $H//ab$.

Fig 8(a) shows the temperature dependency of real ($M'$) and imaginary ($M''$) parts of magnetic $AC$ susceptibility of $Fe_{0.97}Co_{0.03}Se_{0.5}Te_{0.5}$ and $Fe_{0.95}Co_{0.05}Se_{0.5}Te_{0.5}$ single in temperature range 20K to 2K in absence of any $DC$ field. It is clearly seen from the figure that diamagnetic signal occurs at 11K and 4K for $Fe_{0.97}Co_{0.03}Se_{0.5}Te_{0.5}$ and $Fe_{0.95}Co_{0.05}Se_{0.5}Te_{0.5}$ single crystals respectively. The sharp decrement in real part ($M'$) below $T_c$ is basically due to the diamagnetism property of superconductor, confirming the $T_c$. It is clear that superconducting transition temperature ($T_c$) of pristine $FeSe_{0.5}Te_{0.5}$ [28] decreases with increase of Co content in $Fe_{1-x}Co_xSe_{0.5}Te_{0.5}$. The superconducting transition temperature ($T_c$) of $Fe_{1-x}Co_xSe_{0.5}Te_{0.5}$ crystals as being seen from transport (Fig.5) and magnetic (Fig.8) corroborates each other. For further analysis, the isothermal magnetization ($M-H$) plots for $Fe_{0.97}Co_{0.03}Se_{0.5}Te_{0.5}$ single crystal at different temperatures of 2K, 20K and 300K are show in Fig 8(b). The $M-H$ plot at 2K clearly shows the evidence of type II superconductivity. Further, wide opening of $M-H$ plot at 2K till 7Tesla (70kOe) magnetic field suggests high upper critical field for the $Fe_{0.97}Co_{0.03}Se_{0.5}Te_{0.5}$ crystal. This is in tune with the magneto resistivity [$(\rho-T)H$] measurements shown in Figs. 6a and b exhibiting very high upper critical fields for the $Fe_{0.97}Co_{0.03}Se_{0.5}Te_{0.5}$ single crystal. The normal state (above $T_c$) $M-H$ plots for $Fe_{0.97}Co_{0.03}Se_{0.5}Te_{0.5}$ single crystal at 20K and 300K are also shown in Fig. 8(a). The normal state $M-H$ results show the signatures of ferromagnetic ($FM$) like ordering in the crystal, although the signal is quite weak. This may be possible either due to the presence of interstitial Fe [18] in pristine unit cell or tiny impurity of $FeO_x$. Getting rid of interstitial Fe or tiny Fe impurities had been elusive in Fe chalcogenides. For evaluation of lower critical field i.e., $H_{c1}$ of $Fe_{0.97}Co_{0.03}Se_{0.5}Te_{0.5}$ single crystal, we measured low field magnetization at small intervals of applied field and resulting $M-H$ plots at different temperatures are shown in Fig 8(c). In superconducting state, the $M-H$ plots low-field parts principally overlap with the Meissner line due to the perfect shielding and the $H_{c1}(T)$ is defined as the point, where the same starts deviating from the perfect Meissner response. Thus obtained values of $H_{c1}(T)$ for $Fe_{0.97}Co_{0.03}Se_{0.5}Te_{0.5}$ single crystal are approximately 400Oe and 200Oe at 2K and 6K respectively. Thus determined, $H_{c1}(T)$ values (Fig. 8c) are fitted by using the formula $H_{c1}(T) = H_{c1}(0)[1 − (T/T_c)^2]$, shown in Fig. 8(d) and the obtained $H_{c1}(0)$ is 440Oe for the studied $Fe_{0.97}Co_{0.03}Se_{0.5}Te_{0.5}$ single crystal in $H//ab$ condition. Surprisingly though the obtained upper critical field [$H_{c2}(0)$] of $Fe_{0.97}Co_{0.03}Se_{0.5}Te_{0.5}$ single crystal is lower than the pristine $FeSe_{0.5}Te_{0.5}$ single crystal [28], the lower critical field [$H_{c1}(0)$] is higher. In this regards, worth mentioning is the fact that though the conventional single band superconductor equations viz. WHH and GL are routinely applied for determination of critical fields in Fe chalcogenides, the same may not be correct as these systems are known to be multi band superconductors. However in absence of any viable theoretical model for Fe based exotic superconductors, both the WHH and GL are commonly used.

For $Fe_{0.97}Co_{0.03}Se_{0.5}Te_{0.5}$ single crystal, at absolute zero temperature, calculated lower critical field [$H_{c1}(0)$] and upper critical field [$H_{c2}(0)$] are found around 440Oe and 100Tesla respectively. The thermodynamic critical field $H_c(0)$ is calculated using lower and upper critical field at absolute zero temperature, i.e. $H_c(0) = (H_{c1}*H_{c2})^{1/2}$, calculated value of $H_c(0)$ is around 21kOe. According to Ginzburg– Landau theory, upper critical field [$H_{c2}(0)$] and thermodynamic critical field [$H_c(0)$] having the relation i.e. $H_{c2}=2^{1/2}\kappa H_c$, where κ is called kappa parameter which is the Ginzburg– Landau parameter, thus calculated κ is found around 33.6 which is far greater than $1/2^{1/2}$, implying type-II superconductivity in $Fe_{0.97}Co_{0.03}Se_{0.5}Te_{0.5}$. The reported value of κ for Fe-chalcogenide superconductors broadly lies within same range [31,32]. Calculated coherence length $\xi(0)$ for $Fe_{0.97}Co_{0.03}Se_{0.5}Te_{0.5}$ is around 18.14Å. Further, penetration depth $\lambda(0)$ can be calculated using coherence length [$\xi(0)$] and kappa parameter (κ) with the relation i.e. $\lambda(0) = \kappa\xi(0)$, which is found to be around 616Å.

For further description upon the $\rho(T)H$ behaviour of the $Fe_{0.97}Co_{0.03}Se_{0.5}Te_{0.5}$ crystal, the thermally activated flux flow (TAFF) plots i.e., Ln$\rho$ verses Temperature at various fields for both $H//c$ and $H//ab$ are shown in Fig 9(a) and 9(b) respectively. According to TAFF theory [33,34], Ln$\rho$ verses Temperature graph in the TAFF region is described with Arrhenius relation [35] that is given by equation Ln$\rho(T,H)$= Ln$\rho_0(H)$-$U_0(H)/k_BT$., where Ln$\rho_0(H)$ is temperature dependent constant, $U_0(H)$ is called TAFF activation energy and $k_B$ is Boltzmann's constant. From the equation it is clearly seen that in TAFF region, Ln$\rho$ vs $1/T$ graph would be linearly fitted and the fitted linear region with magnetic fields is shown in Fig 9(a) and Fig 9(b) for $H//c$ and $H//ab$ respectively. All the linearly fitted extrapolated lines are intercepted at the same temperature i.e., one being nearly equal to the superconducting transition temperature ($T_c$) which comes around 11.5K and 11.1K for $H//c$ and $H//ab$ respectively. Resistivity broadening with magnetic field is known to be due to thermally assisted flux motion [36]. Seemingly, the resistivity broadening in $Fe_{0.97}Co_{0.03}Se_{0.5}Te_{0.5}$ single crystal is similar to $FeSe_{1-x}Te_x$ based superconductor with increasing magnetic field [37].

Thermally Activation energy of $Fe_{0.97}Co_{0.03}Se_{0.5}Te_{0.5}$ single crystal is calculated for different magnetic fields from range 1Tesla to 14Tesla for both the direction. Fig 9(c) shows the magnetic fields dependency of thermally activation energy for both the direction. As the activation energy is different for both the directions, superconducting anisotropy effect is clearly evident in the studied crystal. Activation energy variation is in wide range with magnetic fields for both directions, this result shows that creep of thermally activated vortices gets affected with increasing magnetic fields. Activation energy varies from 43meV to 12meV with the range of magnetic field from 1Tesla to 12Tesla for *H//c* direction and 36meV to 11meV with the range of magnetic field from 2Tesla to 14Tesla for *H//ab* direction. This activation energy is smaller than the activation energy of $FeSe_{0.5}Te_{0.5}$ single crystal [28]. Magnetic field dependency of thermally activation energy follows power law i.e., $U_0(H)=K\times H^\alpha$ for both direction, where $U_0$ is activation energy, $H$ is field, $K$ is constant and $\alpha$ is field dependent constant. For *H//c* direction, field dependent constant $\alpha=0.34$ for lower field i.e., up to 4Tesla and $\alpha=0.84$ for *H*>6Tesla. While for *H//ab*, $\alpha=0.37$ for field up to 6Tesla and $\alpha=0.98$ for field greater than 6Tesla. For both direction in low magnetic fields, the weak power law of $U_0(H)$ show single vortex pinning in the studied $Fe_{0.97}Co_{0.03}Se_{0.5}Te_{0.5}$ single crystal [38,39]. The thermally activated flux flow (TAFF) behaviour of $Fe_{0.97}Co_{0.03}Se_{0.5}Te_{0.5}$ single crystal is similar to that as reported earlier by us for pristine $FeSe_{0.5}Te_{0.5}$ single crystal [28].

**Conclusion**

We have successfully grown $Fe_{1-x}Co_xSe_{0.5}Te_{0.5}$ (x=0.0 to 0.10) single crystals series with flux free method in a simple programmable automated furnace. From the transport properties it is clearly seen that superconducting transition temperature suppress with the increases in Co concentration at Fe site in $Fe_{1-x}Co_xSe_{0.5}Te_{0.5}$ (x=0.0 to 0.10) single crystal series. SEM images shows the layered by layered morphology at high magnification factor and EDAX results shows the elemental composition is near to stoichiometric ratio. Anisotropy effect of $Fe_{0.97}Co_{0.03}Se_{0.5}Te_{0.5}$ sample clearly shows the single crystalline property up to magnetic fields of 14Tesla. Various superconductivity characteristic parameters including critical fields are given for Co doped superconducting $Fe_{1-x}Co_xSe_{0.5}Te_{0.5}$ single crystals.

**Acknowledgement**

Authors would like to thank their Director NPL India for his keen interest in the present work. This work is financially supported by DAE-SRC outstanding investigator award scheme on search for new superconductors. P. K. Maheshwari thanks CSIR, India for research fellowship and AcSIR-NPL for Ph.D. registration.

**REFERENCES**

1. Y. Kamihara, H. Hiramatsu, M. Hirano, R. Kawamura, H. Yanagi, T. Kamiya, and H. Hosono, J. Am. Chem. Soc. **128,** 10012 (2006).
2. Y. Kamihara, T. Watanabe, M. Hirano, and H. Hosono, J. Am. Chem. Soc. **130,** 3296 (2008).
3. Ren Zhi-An, Lu Wei, Yang Jie, Yi Wei, Shen Xiao-Li, Zheng-Cai, Che Guang-Can, Dong Xiao-Li, Sun Li-Ling, Zhou Fang and Zhao Zhong-Xian, Chin. Phys. Lett. **25**, 2215 (2008).
4. X. H. Chen, T. Wu, G. Wu, R. H. Liu, H. Chen and D. F. Fang, Nature **453**, 761 (2008).
5. F. C. Hsu, J. Y. Luo, K. W. Yeh, T. K. Chen, T. W. Huang, P. M. Wu, Y. C. Lee, Y. L. Huang, Y. Y. Chu, D. C. Yan and M. K. Wu, PNAS **105,** 14262 (2008).
6. K. W. Yeh, T. W. Huang, Y. L. Huang, T. K. Chen, F. C. Hsu, P. M. Wu, Y. C. Lee, Y. Y. Chu, C. L. Chen, and J. Y. Luo, Euro. Phys. Lett. **84**, 37002 (2008)
7. J. G. Bednorz and K. A. Muller, Z. Phys. B **64**, 189 (1986).
8. M.K. Wu, J.R. Ashburn, C.J. Torng, P.H. Hor, R.L. Meng, L. Gao, Z.J. Huang, Y.Q. Wang and C.W. Chu, Phys. Rev. Lett. **58**, 908 (1987)
9. J. Bardeen, L. Cooper and J.R. Schriffer, Phys. Rev. B **8**, 1178 (1957).
10. M. H. Fang, H. M. Pham, B. Qian, T. J. Liu, E. K. Vehstedt, Y. Liu, L. Spinu, and Z. Q. Mao, Phys. Rev. B **78**, 224503 (2008).
11. F. C. Hsu, J. Y. Luo, K. W. Yeh, T. K. Chen, T. W. Huang, P. M. Wu, Y. C. Lee, Y. L. Huang, Y. Y. Chu, D. C. Yan and M. K. Wu, PNAS **105,** 14262 (2008).
12. Y. Mizuguchi, F. Tomioka. S. Tsuda, T. Yamaguchi, and Y. Takano, Appl. Phys. Lett. **93**, 152505 (2008).
13. D. Braithwaite, B. Salce, G. Lapertot, F. Bourdarot, C. Martin, D. Aoki, and M. Hanfland, J. Phys. Cond. Matt. **21**, 232202 (2009).
14. S. Masaki, H. Kotegawa, Y. Hara, H. Tou, K. Murata, Y. Mizuguchi, and Y. Takano, J. Phys. Soc. Jpn. **78**, 063704 (2009)
15. A M Zhang, T L Xia, L R Kong, J H Xiao and Q M Zhang, J. Phys.: Condens. Matter **22**, 245701 (2010)
16. R. Shipra, H. Takeya, K. Hirata and A. Sundaresan, Physica C **470**, 528 (2010)
17. D J Gawryluk1, J Fink-Finowicki1, A Wi´sniewski1, R Pu´zniak1, V Domukhovski1, R Diduszko1,2, M Kozłowski1,2 and M Berkowski, Supercond. Sci. Technol. **24,** 065011 (2011)

18. Anuj Kumar, R.P. Tandon and V.P.S. Awana, IEEE trans. mag. **48**, 4239 (2012).
19. Z. T. Zhang, Z. R. Yang, L. Li, C. J. Zhang, L. Pi, S. Tan, Y. H. Zhang, J. Appl. Phys. **109**, 07E113 (2011).
20. B. H. Mok, S. M. Rao, M. C. Ling, K. J. Wang, C. T. Ke, P. M. Wu, C. L. Chen, F. C. Hsu, T. W. Huang, J. Y. Luo, D. C. Yan, K. W. Ye, T. B. Wu, A. M. Chang and M. K. Wu, Cryst. Growth Design. **9**, 3260 (2009).
21. U. Patel, J. Hua, S. H. Yu, S. Avci, Z. L. Xiao, H. Claus, J. Schlueter, V. V. Vlasko-Vlasov, U. Welp and W. K. Kwok, Appl. Phys. Lett. **94**, 082508 (2009).
22. T. J. Liu, J. Hu, B. Qian, D. Fobes, Z. Q. Mao, W. Bao, M. Reehuis, S. A. J. Kimber, K. Prokes, S. Matas, D. N. Argyriou, A. Hiess, A. Rotaru, H. Pham, L. Spinu, Y. Qiu, V. Thampy, A. T. Savici, J. A. Rodriguez and C. Broholm, Nat. Mater. **9**, 718 (2010).
23. S. I. Vedeneev, B. A. Piot, D. K. Maude, and A. V. Sadakov, Phys. Rev. B **87**, 134512 (2013).
24. J. Wen, G. Xu, G. Gu, J.M. Tranquada, and R.J. Birgeneau, Rep. Prog. Phys. **74**, 124503 (2011).
25. D. P. Chen and C.T. Lin, Sup. Sci. & Tech. **27**, 103002 (2014).
26. B.C. Sales, A.S. Sefat, M.A. McGuire, R.Y. Jin, D. Mandrus, and Y. Mozharivskyj, Phys. Rev. B. **79**, 094521 (2009).
27. M. Ma, D. Yuan, Y. Wu, Z. Huaxue, X. Dong, and F. Zhou, Sup. Sci. & Tech. **27**, 122001 (2014)
28. P. K. Maheshwari, Rajveer Jha, B. Gahtori and V. P. S. Awana, AIP Advances **5**, 097112 (2015).
29. P.K. Maheshwari, L.M. Joshi, Bhasker Gahtori, A.K. Srivastava, Anurag Gupta. S.P. Patnaik and V.P.S. Awana, Mater. Res. Express **3**, 076002 (2016).
30. P.K. Maheshwari, Bhasker Gahtori and V.P.S. Awana, J. Sup. & Novel Mag. **29**, 2473 (2016).
31. Hongyan Yu, Ming Zuo, Lei Zhang, Shun Tan, Changjin Zhang, and Yuheng Zhang, J. Am. Chem. Soc. **135**, 12987 (2013).
32. Hechang Lei, Rongwei Hu and C. Petrovic, Phys. Rev. B **84**, 014520 (2011).
33. T. T. M. Palstra, B. Batlogg, L. F. Schneemeyer, and J. V. Waszczak, Phys. Rev. Lett. **61**, 1662 (1988).
34. G. Blatter, M. V. Feigelman, V. B. Geshkenbein, A. I. Larkin, and V. M. Vinokur, Rev. Mod. Phys. **66**, 1125 (1994).

35. J. Jaroszynski, F. Hunte, L. Balicas, Y. -J. Jo, I. Raicevic, A. Gurevich, D. C. Larbalestier, F. F. Balakirev, L. Fang, P. Cheng, Y. Jia, and H .H. Wen, Phys. Rev. B **78**, 174523 (2008).

36. M. Shahbazi, X. L. Wang, C. Shekhar, O. N. Srivastava, and S. X. Dou, Sup. Sci. & Tech **23,** 105008 (2010).

37. K. W. Yeh, H. C. Hsu, T. W. Huang, P. M. Wu, Y. L. Yang, T. K. Chen, J. Y. Luo, and M. K.Wu, J. Phys. Soc. Jpn. **77**, 19 (2008).

38. G. Blatter, M. V. Feigelman, V. B. Geshkenbein, A. I. Larkin, and V. M. Vinokur, Rev. Mod. Phys. **66**, 1125(1994).

39. H. Lei, R. Hu and C. Petrovic, Physical Review B **84**, 014520 (2011).

**Table 1:** $Fe_{1-x}Co_xSe_{0.5}Te_{0.5}$ (x=0.0 to 0.10) single crystals lattice parameters and coordinate positions using Rietveld refinement.

|  | **x= 0.00** | **x=0.03** | **x=0.05** | **x=0.10** |
|---|---|---|---|---|
| $a=b$ (Å) | 3.801(2) | 3.792(4) | 3.788(3) | 3.780(2) |
| $c$ (Å) | 5.998(4) | 5.9640(3) | 5.950(3) | 5.935(3) |
| V(Å3) | 86.656(3) | 85.757(3) | 85.376(2) | 84.801(2) |
| Fe | (3/4,1/4,0) | (3/4,1/4,0) | (3/4,1/4,0) | (3/4,1/4,0) |
| Co | - | (3/4,1/4,0) | (3/4,1/4,0) | (3/4,1/4,0) |
| Se | (1/4,1/4,0.286) | (1/4,1/4,0.256) | (1/4,1/4,0.249) | (1/4,1/4,0.246) |
| Te | (1/4,1/4,0.286) | (1/4,1/4,0.256) | (1/4,1/4,0.249) | (1/4,1/4,0.246) |

**FIGURE CAPTIONS**

**Figure 1:** Photograph of $Fe_{1-x}Co_xSe_{0.5}Te_{0.5}$ (x=0.0 to 0.10) single crystals.

**Figure 2:** Selected SEM images and EDAX results (a) $Fe_{0.97}Co_{0.03}Se_{0.5}Te_{0.5}$ single crystal SEM image at 2µm (b) $Fe_{0.90}Co_{0.10}Se_{0.5}Te_{0.5}$ single crystal SEM image at 20µm (c) EDAX Quantitative analysis of $Fe_{0.97}Co_{0.03}Se_{0.5}Te_{0.5}$ single crystal of selected area (d) EDAX Quantitative analysis of $Fe_{0.90}Co_{0.10}Se_{0.5}Te_{0.5}$ single crystal of selected area

**Figure 3:** Single crystal surface XRD for $Fe_{1-x}Co_xSe_{0.5}Te_{0.5}$ (x=0.0 to 0.10) at room temperature.

**Figure 4:** Powder XRD patterns for crushed powders of $Fe_{1-x}Co_xSe_{0.5}Te_{0.5}$ (x=0.0, 0.03, 0.05 and 0.10) single crystals at room temperature. *Inset* view is the expanded [003] plane view for the same.

**Figure 5:** Normalised resistivity ($\rho/\rho_{300}$) versus temperature plots for $Fe_{1-x}Co_xSe_{0.5}Te_{0.5}$ (x=0.0 to 0.10) single crystals in temperature range of 2 to 300K. Inset view is zoomed view of same in temperature range of 2K to 20K.

**Figure 6:** Temperature dependency resistivity $\rho(T)$ under various magnetic fields up to 14Tesla for (a) *H//c* plane and (b) *H//ab* plane for $Fe_{0.97}Co_{0.03}Se_{0.5}Te_{0.5}$ single crystal.

**Figure 7:** Upper critical field ($H_{c2}$) calculated from the $\rho(T)H$ data with 90%, 50% and 10% $\rho_n$ criteria for (a) *H//c* plane and (b) *H//ab* plane for $Fe_{0.97}Co_{0.03}Se_{0.5}Te_{0.5}$ single crystal.

**Figure 8:** (a) *AC* magnetic susceptibility at 333Hz and 5Oe amplitude for $Fe_{0.97}Co_{0.03}Se_{0.5}Te_{0.5}$ and $Fe_{0.95}Co_{0.05}Se_{0.5}Te_{0.5}$ single crystal. (b) Isothermal *M-H* curve for $Fe_{0.97}Co_{0.03}Se_{0.5}Te_{0.5}$ at 2K, 20K and 300K temperatures. (c) Low field *M-H* curve at temperature range for 2K to 6K of $Fe_{0.97}Co_{0.03}Se_{0.5}Te_{0.5}$ single crystal. (d) $H_{c1}(T)$ vs *T*, fitted solid line is fitting to $H_{c1}(T) = H_{c1}(0)[1-(T/Tc)^2]$ for $Fe_{0.97}Co_{0.03}Se_{0.5}Te_{0.5}$ single crystal.

**Figure 9:** $\ln\rho(T,H)$ vs $1/T$ for different magnetic fields (a) *H//c* and (b) *H//ab* plane for $Fe_{0.97}Co_{0.03}Se_{0.5}Te_{0.5}$ single crystal corresponding fitted solid line of Arrhenius relation. (c) Thermally Activation energy $U_o(H)$ with solid lines fitting of $U_o(H) \sim H^\alpha$ for different magnetic field.

Fig 1

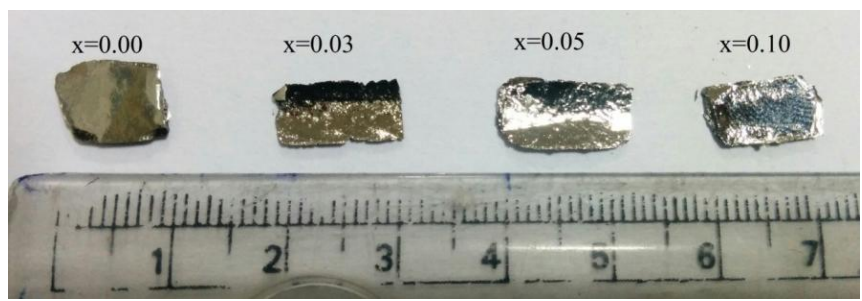

Fig 2

(a) (b)

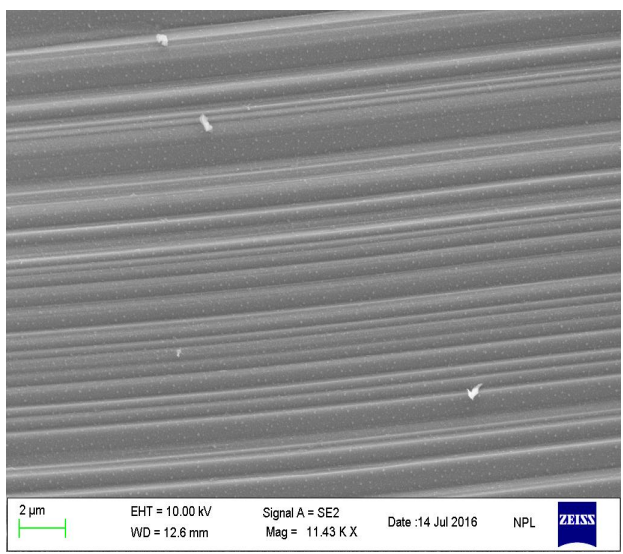
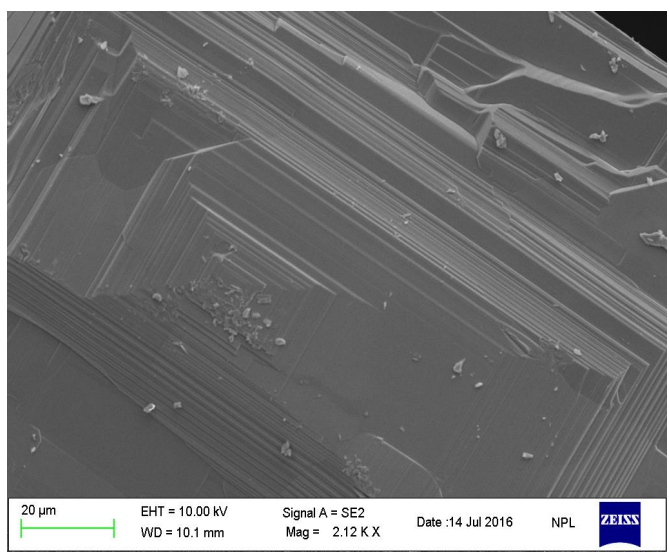

(c) (d)

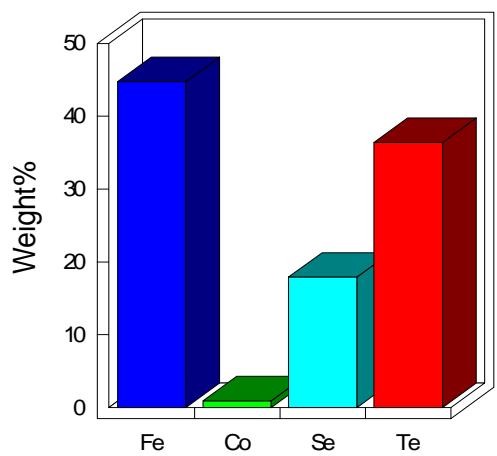
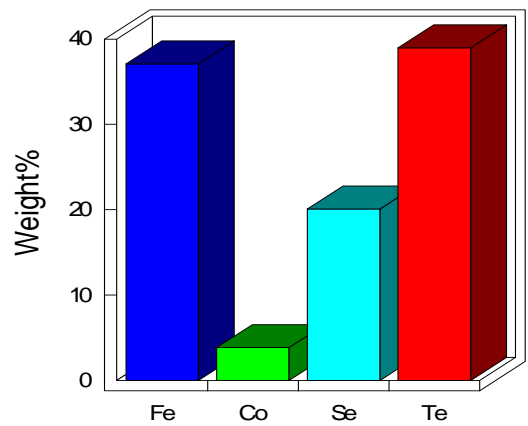

Figure 3

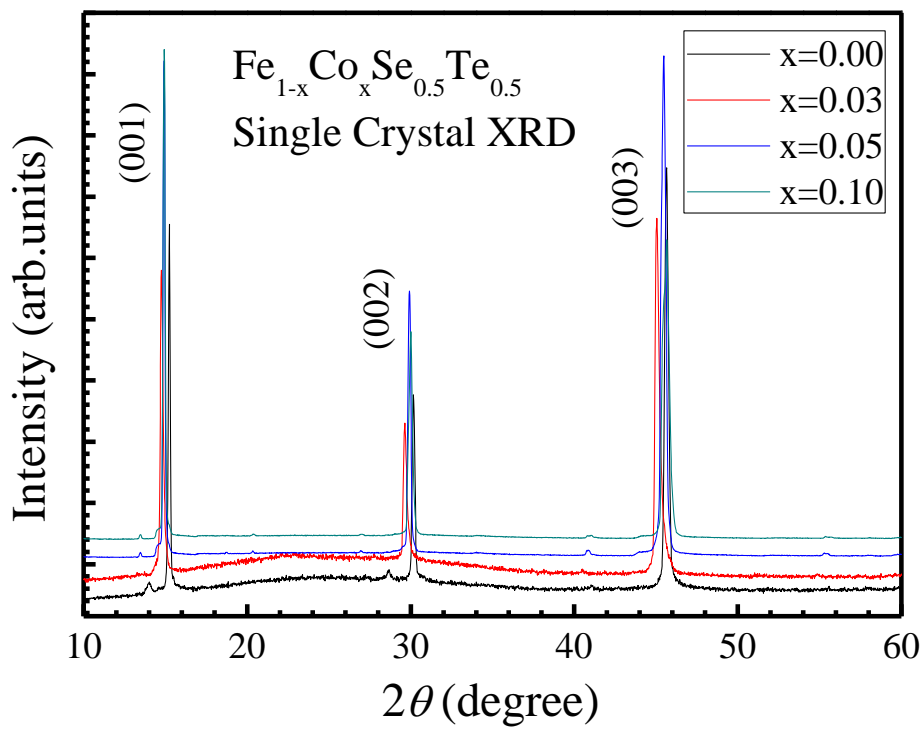

Figure 4

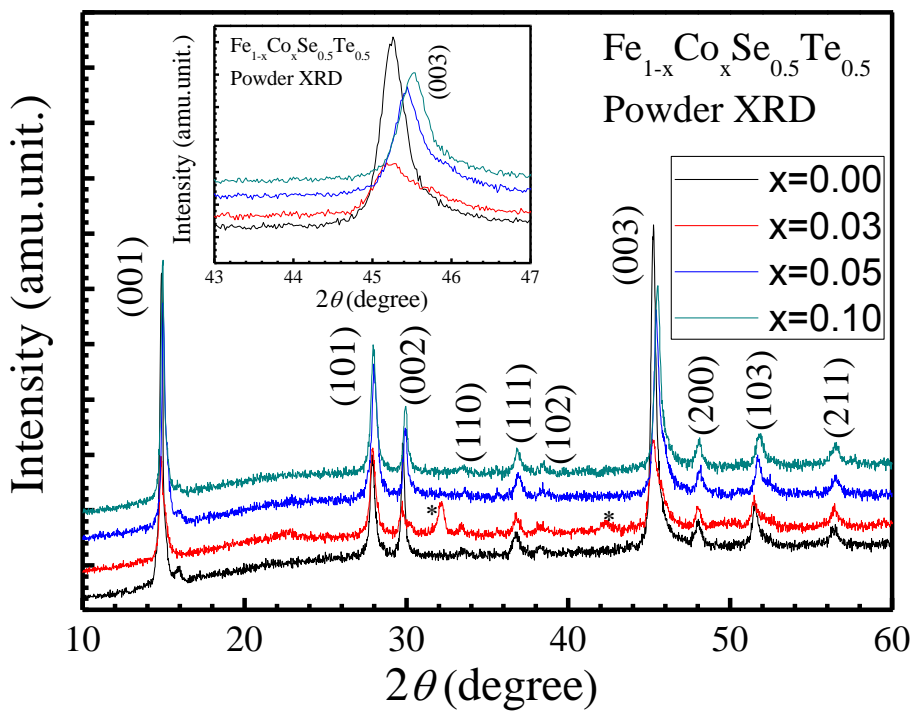

Figure 5

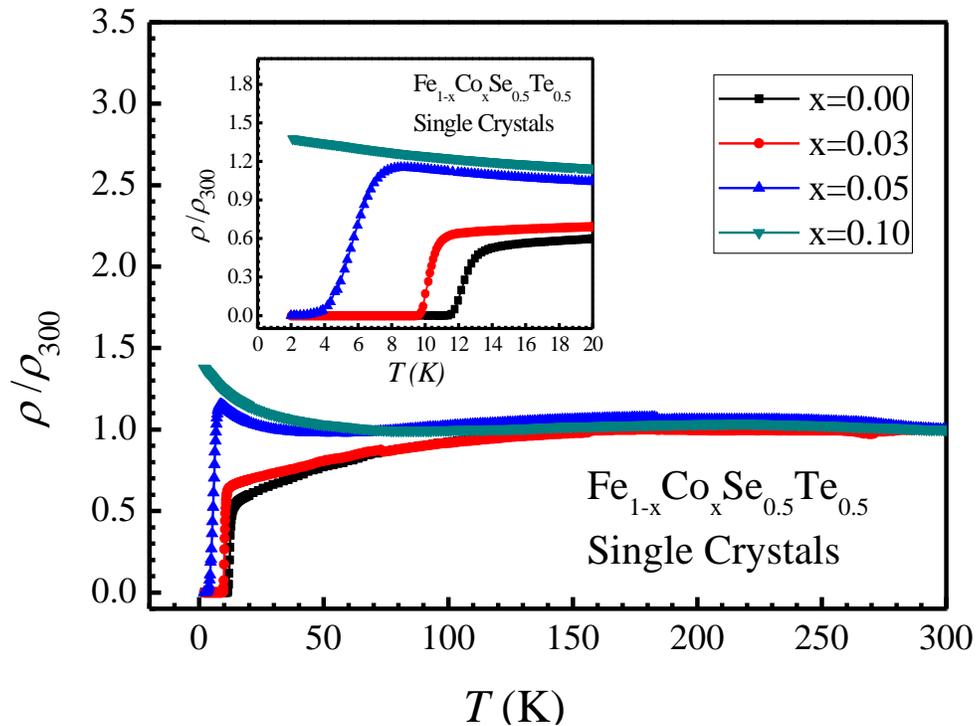

Figure 6 (a)

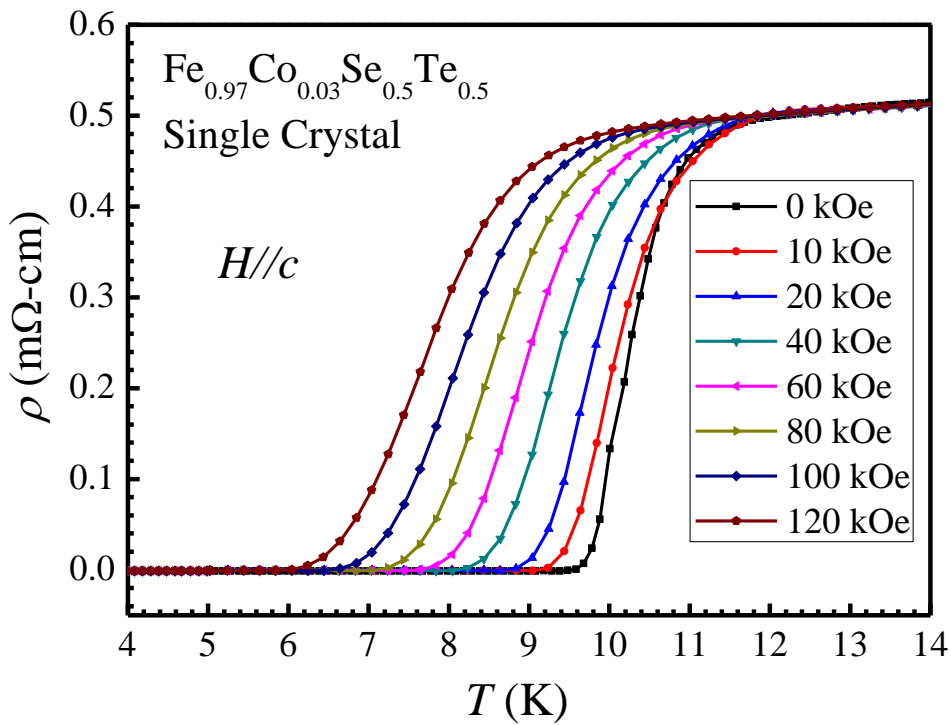

Figure 6 (b)

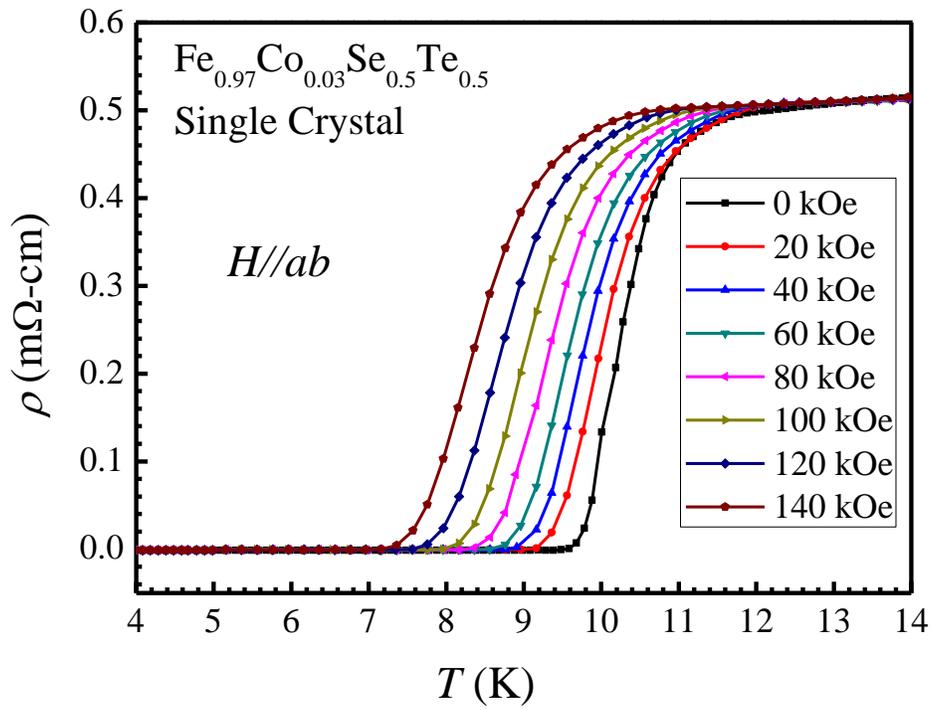

Figure 7 (a)

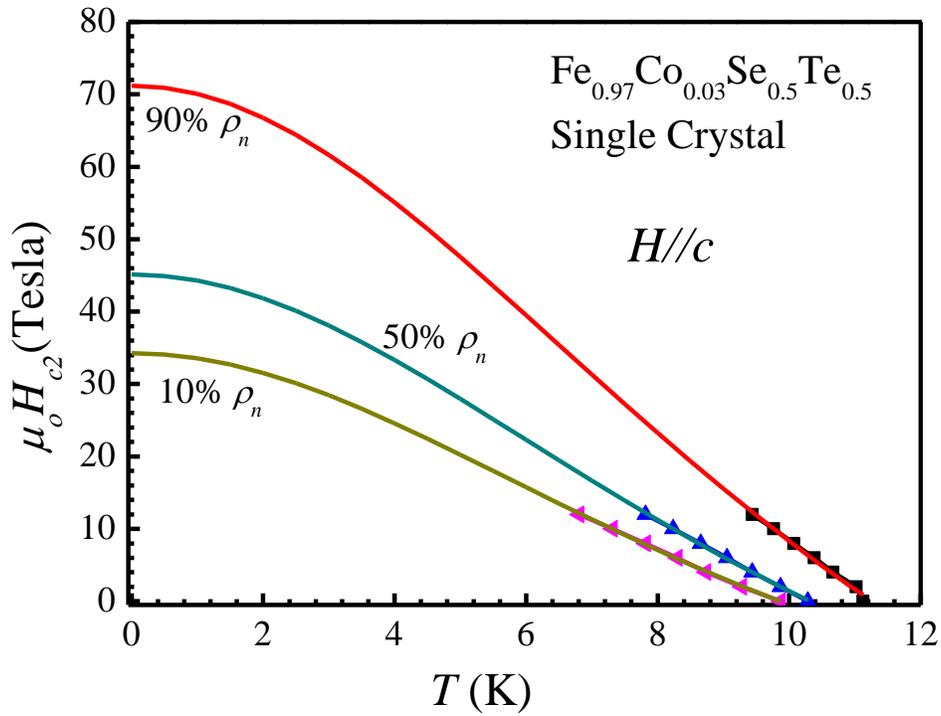

Figure 7 (b)

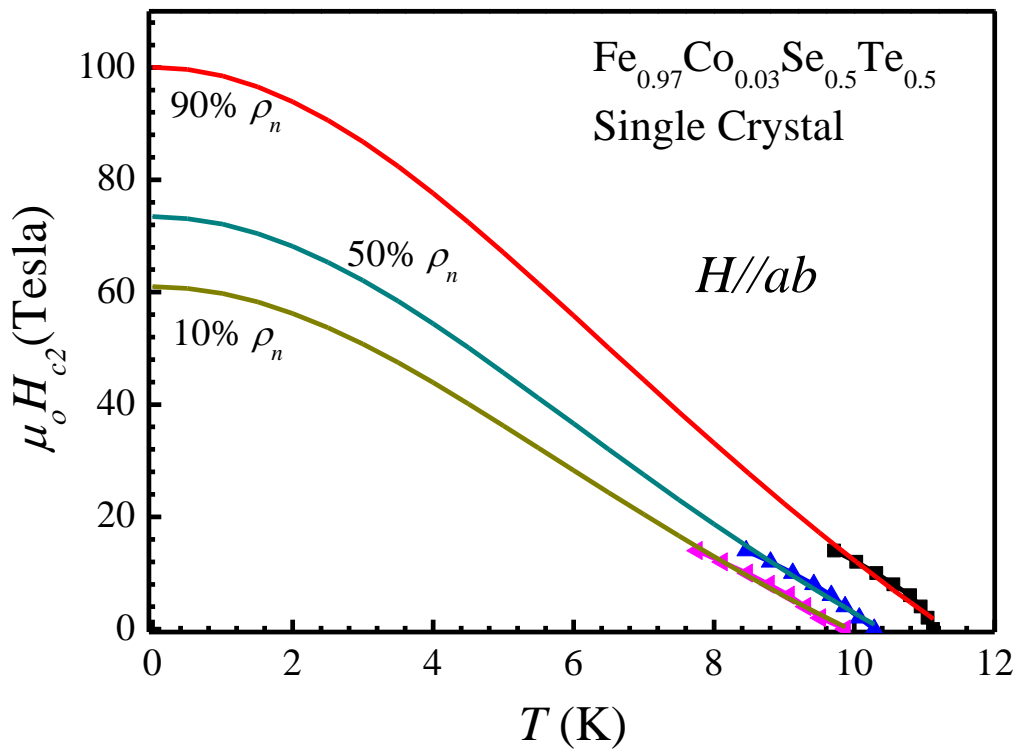

Figure 8 (a)

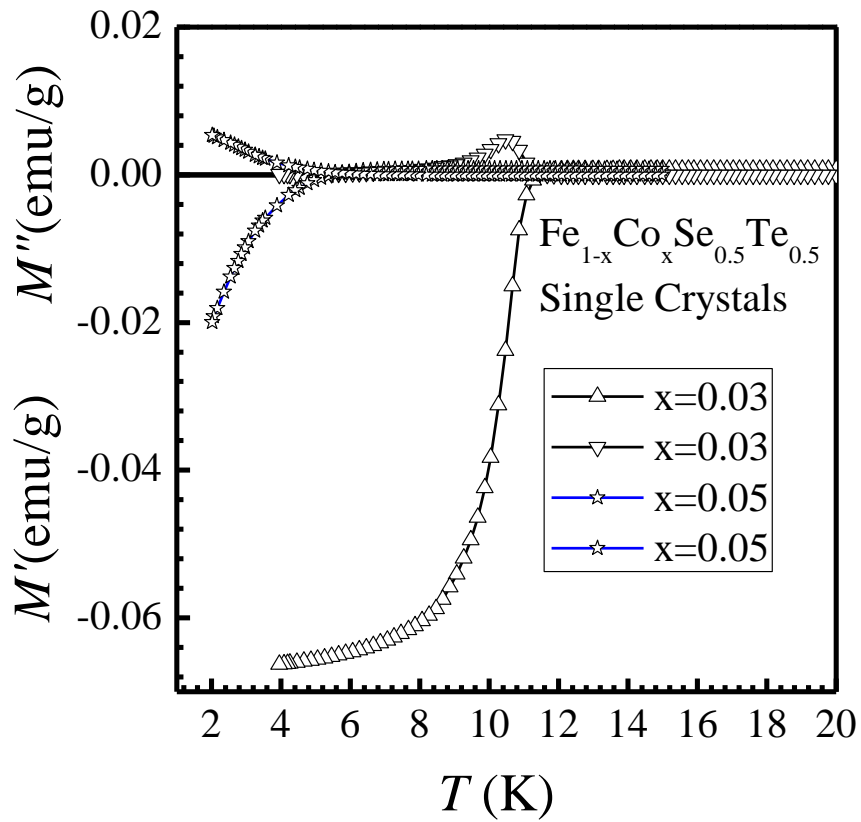

Figure 8 (b)

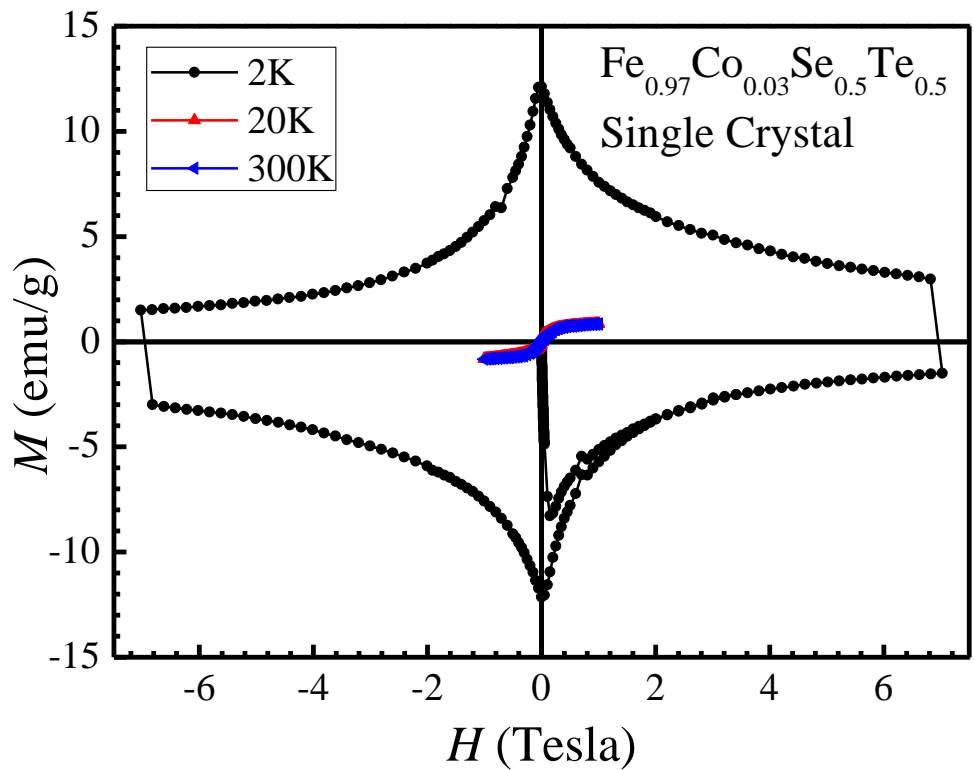

Figure 8 (c)

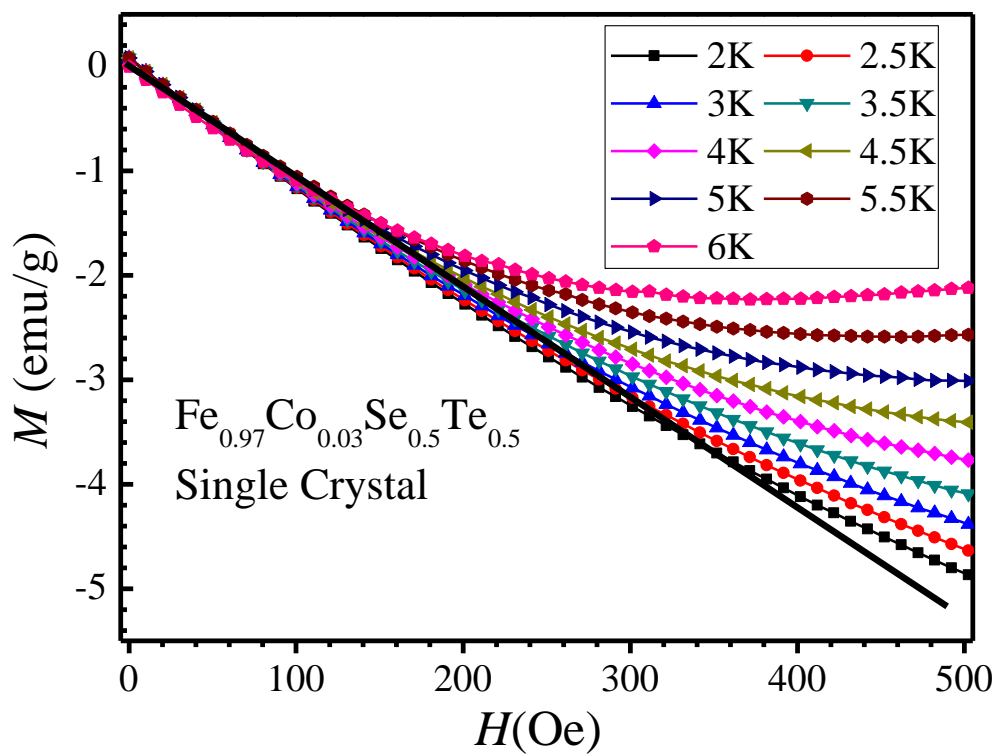

Figure 8(d)

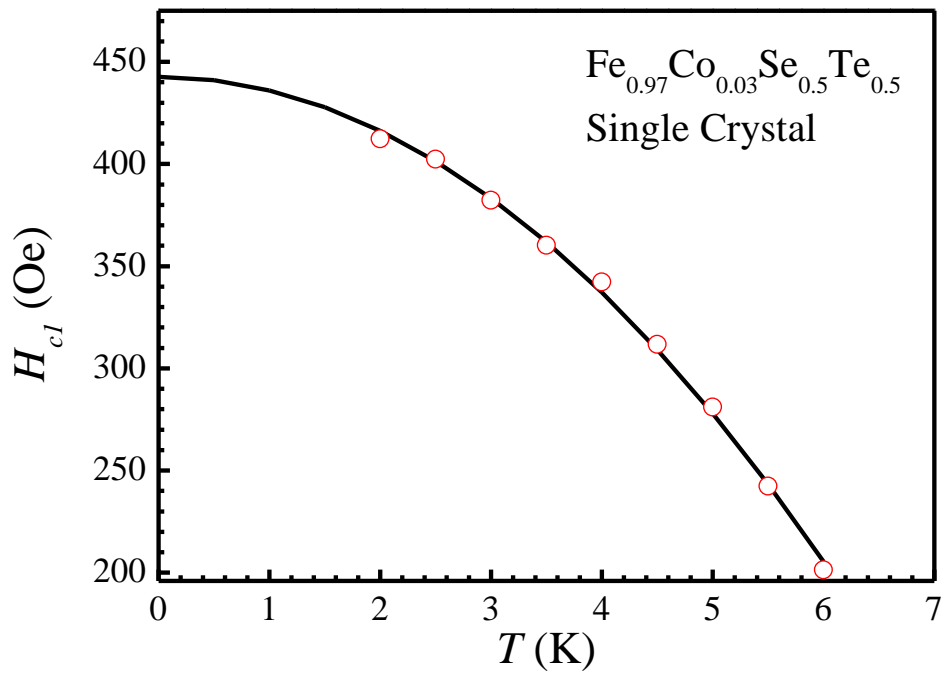

Figure 9 (a)

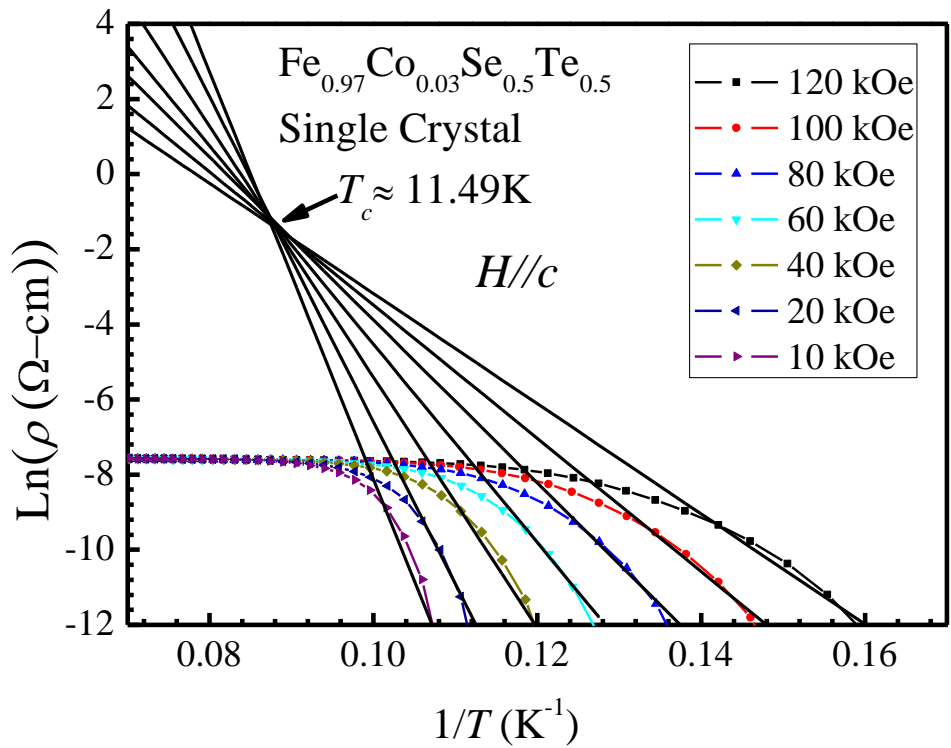

Figure 9 (b)

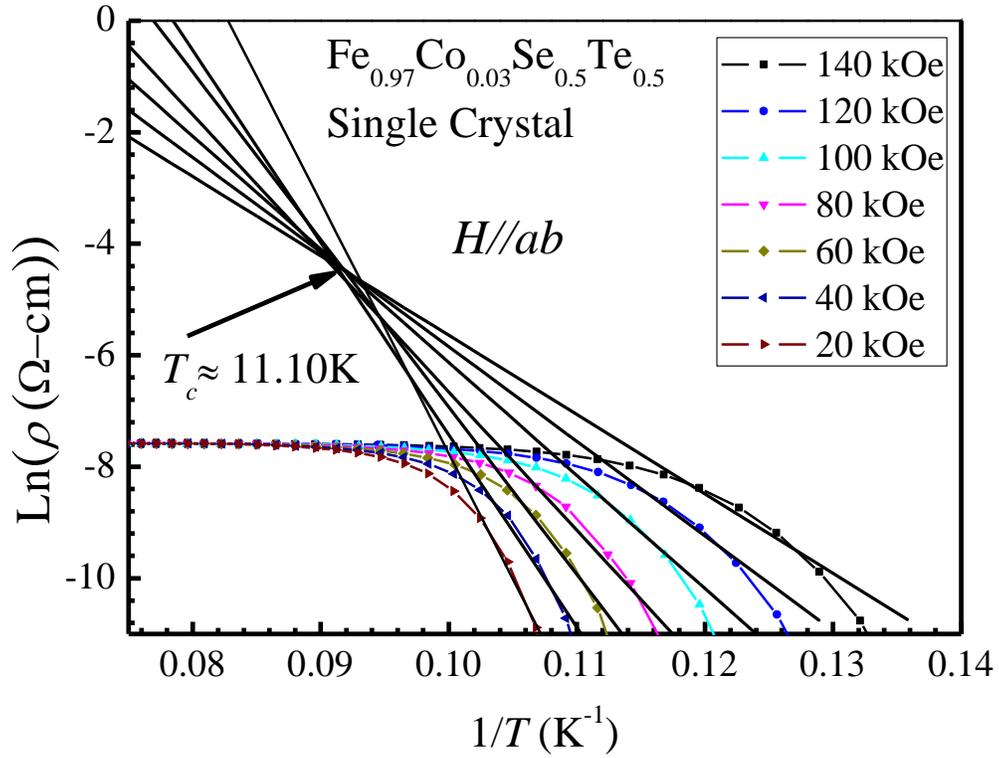

Figure 9(c)

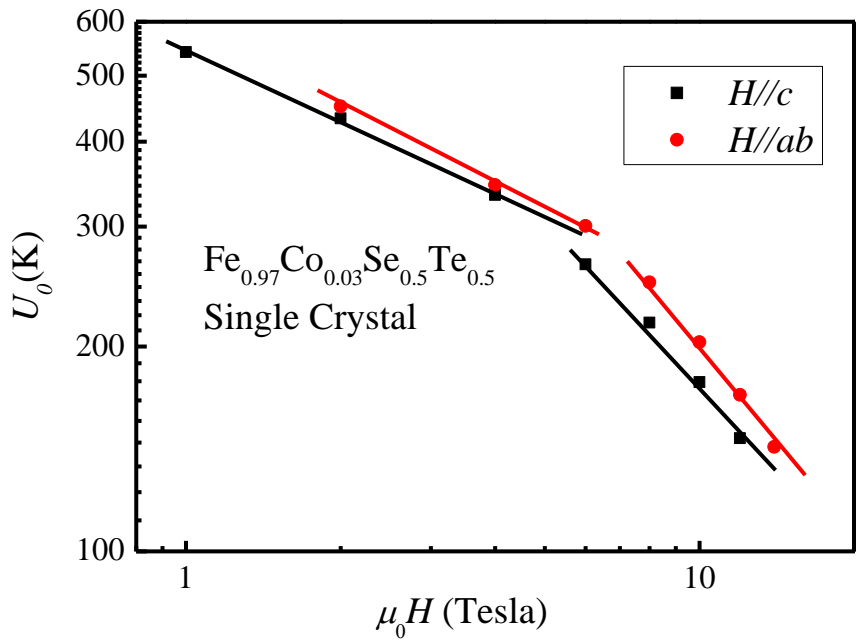